# Creation of a novel inverted charge density wave state

Yingchao Zhang[1]†, Xun Shi[1]*†, Mengxue Guan[2]†, Wenjing You[1], Yigui Zhong[1,2], Tika R. Kafle[1], Yaobo Huang[3], Hong Ding[2], Michael Bauer[4], Kai Rossnagel[4], Sheng Meng[2], Henry C. Kapteyn[1], and Margaret M. Murnane[1]

[1]Department of Physics and JILA, University of Colorado and NIST, Boulder, CO 80309, USA

[2]Beijing National Laboratory for Condensed Matter Physics and Institute of Physics, Chinese Academy of Sciences, Beijing 100190, China

[3]Shanghai Synchrotron Radiation Facility, Shanghai Advanced Research Institute, Chinese Academy of Sciences, Shanghai 201204, China

[4]Institute of Experimental and Applied Physics, Kiel University, D-24098 Kiel, Germany

*Corresponding author. Email: Xun.Shi@colorado.edu

†These authors contributed equally to this work.

Charge density wave (CDW) order is an emergent quantum phase that is characterized by a periodic lattice distortion and charge density modulation, often present near superconducting transitions. Here we uncover a novel inverted CDW state by using a femtosecond laser to coherently over-drive the unique star-of-David lattice distortion in 1$T$-TaSe$_2$. We track the signature of this novel CDW state using time- and angle-resolved photoemission spectroscopy and time-dependent density functional theory, and validate that it is associated with a unique lattice and charge arrangement never before realized. The dynamic electronic structure further reveals its novel properties, that are characterized by an increased density of states near the Fermi level, high metallicity, and altered electron-phonon couplings. Our results demonstrate how ultrafast lasers can be used to create unique states in materials, by manipulating charge-lattice orders and couplings.

The correlated interactions of electrons and atoms in crystalline materials, including the electron-phonon coupling [1,2], play a crucial role in stabilizing emergent quantum phases such as charge density wave (CDW) order and superconductivity. Despite decades of research, the underlying physics of these phases is still elusive. Recently, first-principle calculations of electron-phonon coupling have made it possible to predict the properties and behaviors of real materials [1]. Techniques such as Raman [3], neutron [4] and X-ray scattering [5] can probe phonon modes in the frequency domain and provide information about the integrated electron-phonon coupling over all electronic bands. For a comprehensive investigation of electron-phonon coupling, one needs to determine the electron-phonon matrix element $g_{mn\nu}(\mathbf{k}, \mathbf{q})$, which measures the coupling strength for the electron scattering from an initial state in band $n$ with momentum $\mathbf{k}$, to a final state in band $m$ with momentum $\mathbf{k} + \mathbf{q}$, via a phonon with mode $\nu$ at wave vector $\mathbf{q}$. However, it remains challenging to experimentally extract momentum-resolved information about specific phonon modes interacting with particular bands.

Ultrafast laser excitation of quantum materials provides unique access to new light-induced states and their dominant couplings [6–13]. When combined with advanced ultrafast spectroscopy techniques, it is now possible to study electron-phonon coupling in the time domain [14–22]. Among them, time- and angle-resolved photoemission spectroscopy (trARPES) combines the direct measurement of electronic structure, with the ability to probe electron-phonon couplings on their intrinsic timescales – from femtoseconds on up. By monitoring either phonon-mediated electron scattering processes [23] or band dynamics in the presence of coherent phonons [24–27], certain phonon modes can be selectively excited or coupled. In particular, ultrafast laser-excited coherent phonons [28] can modulate the electronic structure in a momentum-dependent way, which can be characterized by band- and mode-projected electron-phonon coupling strengths [29,30]. Thus, how the electronic order responds to coherent phonons provides unique opportunities for studying and manipulating electron-phonon coupling in quantum materials, and importantly, map the light-enhanced phase diagram.

In this work, we uncover a novel light-induced inverted CDW state in $1T$-TaSe$_2$ that has not been observed previously, and validate its nature through a comparison of multiple experimental signatures with theory. This material has a unique star-of-David periodic lattice distortion that can be coherently excited by an ultrafast laser pulse. This launches a coherent CDW amplitude mode at ~ 2 THz. Using trARPES, we can monitor the dynamic electronic structure and CDW order, which is encoded in the Ta 5$d$ band position. For sufficiently strong laser excitation, the coherent amplitude mode drives the material from the usual CDW state, through the normal state, before entering a new inverted CDW state. In particular, this inverted CDW state is associated with an overshoot of the star-of-David periodic lattice distortion, that supports a new CDW state exhibiting a high metallic character (Fig. 1). These results are confirmed by simulations based on time-dependent density functional theory (TDDFT), which allow us to investigate band- and mode-projected electron-phonon coupling across different states.

A great advantage of trARPES is that we can measure the band position, dispersion and folding, as well as the CDW gap, providing multiple observations to compare with theory. We find that the Ta 5$d$ bandwidth (*i.e.* the energy range of the full dispersive band) along the Γ-M direction increases as the material transforms from the usual CDW state into the normal state, before decreasing again as the material enters an inverted CDW state. In contrast, the band position shifts upwards monotonically. This evolution of bandwidth *vs.* band position originates from the momentum-dependent electron-phonon coupling between the coherent amplitude mode and the Ta 5$d$ band, and reveals an inversion of deformation potential gradient in the momentum space when the material enters the new inverted CDW state — which is closely related to a change in the electron hopping interaction.

1$T$-TaSe$_2$ is a typical two-dimensional (2D) CDW material. In the CDW state (below 470 K), the lattice reconstructs into a $\sqrt{13} \times \sqrt{13}$ star-of-David supercell, illustrated in the upper left panel of Fig. 1 [31]. The lattice reconstruction leads to the band folding, giving rise to multiple band crossings and gap opening. Figure S1 shows the static ARPES measurements of the band structure at 300 K. In the Ta 5$d$ band near the Fermi level ($E_F$), two pronounced CDW gaps are observed along the Γ-M and M-K directions, respectively.

Our trARPES measurements employ a 1.6 eV pump pulse, followed by a 22.4 eV probe pulse produced from high harmonic generation (HHG). After laser excitation of 1$T$-TaSe$_2$, the electrons thermalize to thousands of Kelvin within ~50 fs. The laser-excited electron redistribution into unoccupied states smears out the charge localization, which defines a new equilibrium position for the periodic lattice distortion, and displacively launches the coherent amplitude mode [11]. By strongly exciting this coherent mode, we can generate a unique and otherwise unreachable periodic lattice distortion (Fig. 1 top right) — that is very relevant to the emergent properties of 1$T$-TaSe$_2$. This process can be dynamically monitored because it is manifested by changes in multiple prominent features in the ARPES spectrum — such as the Ta 5$d$ band position and dispersion, the CDW gap, and the band folding (Figs. 2(a)–2(c)). The blue curves in Figs. 2(e)–2(f) represent the time evolution of the band shift near the Γ point with respect to the normal state position, for two representative pump laser fluences. Additional results can be found in Fig. S2.

When the laser excitation is sufficiently strong to completely smear out the charge localization, the equilibrium point of the amplitude mode oscillation is at zero lattice distortion. However, as shown in Figs. 2(f) and S2, transient overshooting of the periodic lattice distortion (as reflected in the band shift) leads to the formation of an inverted CDW state, which mainly persists within the first cycle of the amplitude mode. As the material is coherently excited from the usual CDW state, to the normal state, and then into the inverted CDW state, the Ta 5$d$ band shifts up towards $E_F$ *monotonically*, making this new state metallic.

To gain deep insight into the coherent electron and lattice dynamics after laser excitation, nonadiabatic molecular dynamic simulations based on TDDFT were carried out. The dynamics of CDW phase under two laser fluences are shown in Fig. 3(a). To quantify the lattice structural changes, the root-mean-square displacement (RMSD$(t)=\sqrt{\mu^2(t)}$) is calculated, where $\mu$ describes the atomic displacement of all atoms relative to the initial CDW structure. We adopt a criterion that the CDW state melts or transforms to a new phase when the RMSD reaches $R_c$ = 0.24 Å, based on the fact that the CDW pattern is distorted by 7% from the normal state [32]. For lower fluence (0.17 mJ/cm$^2$), the RMSD increases slowly but stays below the melting threshold $R_c$. However, for higher fluence (0.61 mJ/cm$^2$), the RMSD value exceeds $R_c$ at 347 fs (close to the normal state), and reaches a maximum of 0.3 Å at 436 fs, accompanying the emergence of a new transient state. The calculated structure in the upper right panel of Fig. 3(a) confirms that it is indeed associated with an overshoot of the periodic lattice distortion (expanded star-of-David). Importantly, the calculated unfolded band structures [33,34] in Fig. 3(b) suggest that the inverted

CDW state is highly-metallic, with larger density of states near $E_F$ than the normal metallic state, in strong contrast to the usual CDW state (Fig. 3(c)).

Note that the CDW patterns of the original and inverted CDW states are geometrically equivalent for many CDW materials such as Cr [35], TiSe$_2$ [36,37], $R$Te$_3$ ($R$: rare-earth element) [38] (lower panel of Fig. 1 as an example). Because of the special $\sqrt{13} \times \sqrt{13}$ star-of-David supercell, the inverted CDW in 1$T$-TaSe$_2$ is significantly distinct from the usual CDW in terms of the lattice order (upper panel of Fig. 1), as well as the electronic structure, which defines it as a truly new state.

To further investigate the properties of this novel inverted CDW state, we monitor the Ta 5$d$ bandwidth (as defined in Fig. 2(d)), which is closely related to the electron-electron correlation and/or the electron-phonon coupling [1,39–42]. Note that the bandwidth is also strongly modified by the band folding and hybridization, and we take this into account. In Fig. 2(a), the two original unreconstructed bands (red curves) can be extracted from the ARPES spectra after removing the hybridization gap using a mean-field approximation (see Supplemental Material). Figures 2(e)–2(f) plot the dynamics of this bandwidth, together with the band shift for two typical laser pump fluences. For lower fluence (0.24 mJ/cm$^2$), the bandwidth and band shift dynamics are similar, both exhibiting coherent oscillations. However, for higher fluence (0.86 mJ/cm$^2$), they deviate on short timescales. Figure 4 plots the bandwidth as a function of band shift over a wide range of fluence, to clearly highlight the different regimes. The bandwidth increases as the material evolves from the usual CDW state to the normal state. However, this trend reverses as the material enters the inverted CDW state. This result is also confirmed by the DFT calculations (Fig. 3(b)).

Such a relation between the bandwidth and the band shift shown in Fig. 4 provides unique insight into mode-projected electron-phonon coupling during the ultrafast phase transitions. The momentum-dependent band shift of the Ta 5$d$ band, which leads to a dynamical bandwidth, can be characterized by the momentum-dependent deformation potential for the specific amplitude mode: $D(\boldsymbol{k}) = \delta\varepsilon(\boldsymbol{k})/\delta u$, where $\delta\varepsilon(\boldsymbol{k})$ is the band shift and $\delta u$ is the variation of phonon coordinate. $D(\boldsymbol{k})$, which determines the electron-phonon coupling strength, plays a pivotal role in structural phase transitions. The band dispersions across the three material states we probe are schematically presented in Fig. 5. The color scale of the shading qualitatively represents the evolution of $D(\boldsymbol{k})$ in the momentum space. Between the usual CDW and normal states, $D(\boldsymbol{k})$ at the Γ point is larger than that at the M point, while the trend is reversed on the inverted CDW side. When the periodic lattice distortion crosses zero, there is a change in both the sign and magnitude of $\partial D(\boldsymbol{k})/\partial \boldsymbol{k}$, as manifested by the opposite and asymmetric slope (bandwidth *vs.* band shift) in Fig. 4. This inversion of deformation potential gradient could be related to the unique expanded David-stars in the inverted CDW state compared to the usual contracted ones (Fig. 1). Such a lattice rearrangement significantly modifies the inter-star, intra-star and interlayer electron hopping interactions, which result in a narrowing of the bandwidth in the inverted CDW state. Future research such as ultrafast diffraction experiments could provide deeper understandings on this fundamental problem.

In summary, we demonstrate the creation of a novel inverted CDW state in the 2D CDW material 1$T$-TaSe$_2$ using trARPES and TDDFT. An intense femtosecond laser pulse can strongly excite the coherent phonon to drive the material into an otherwise unreachable state with unique order. Time-resolved ARPES can then monitor the dynamic electronic response to this specific mode in a band- and momentum-resolved manner. This allows us to investigate the electronic structure and electron-phonon couplings in many unique lattice configurations, and further stimulate advanced theory.

**Acknowledgements:** We thank Xianxin Wu and Daniel T. Larson for useful discussions. M.M. and H.K. gratefully acknowledge support from the National Science Foundation through the JILA Physics Frontiers Center PHY-1125844 and a Gordon and Betty Moore Foundation EPiQS Award GBMF4538. H.D. gratefully acknowledge support from the National Natural Science Foundation of China (11888101 and 11674371), and Ministry of Science and Technology of China (2016YFA0401000).

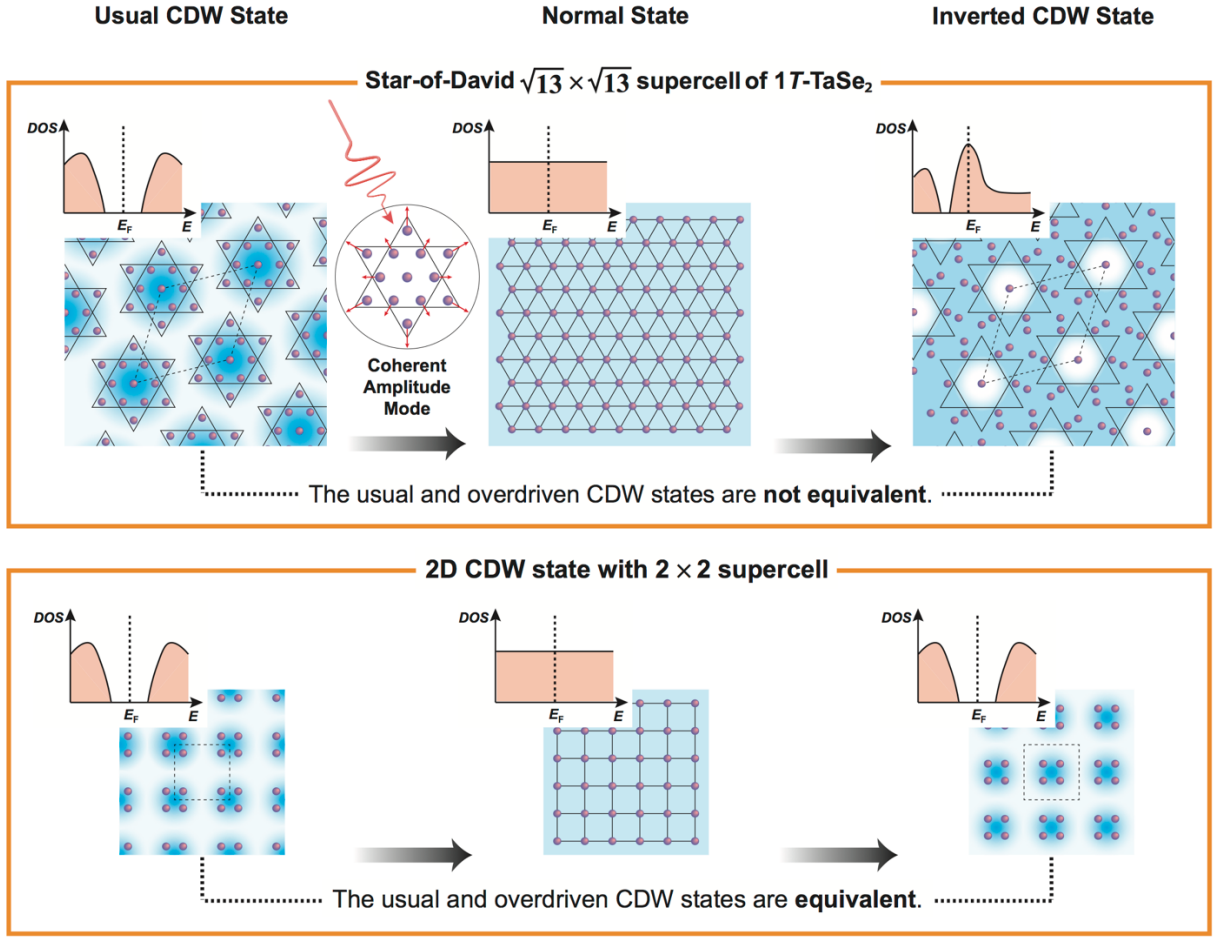

FIG. 1. Generation of an inverted CDW state via ultrafast laser excitation of the coherent amplitude mode. The upper panel illustrates the evolution from the usual CDW state to the normal state and then to a new inverted CDW state in $1T$-TaSe$_2$. The blue shading represents the spatial electron density, while the purple circles represent in-plane Ta atoms with exaggerated lattice distortion. Because of the unique $\sqrt{13} \times \sqrt{13}$ supercell, the inverted CDW state is distinct from the usual one and shows high-metallicity, as shown in the schematic density of states (top inset of each panel). The lower panel shows an example case of a $2 \times 2$ supercell for comparison, where the usual and inverted CDW states are equivalent, only differing by a spatial phase shift of $\pi$.

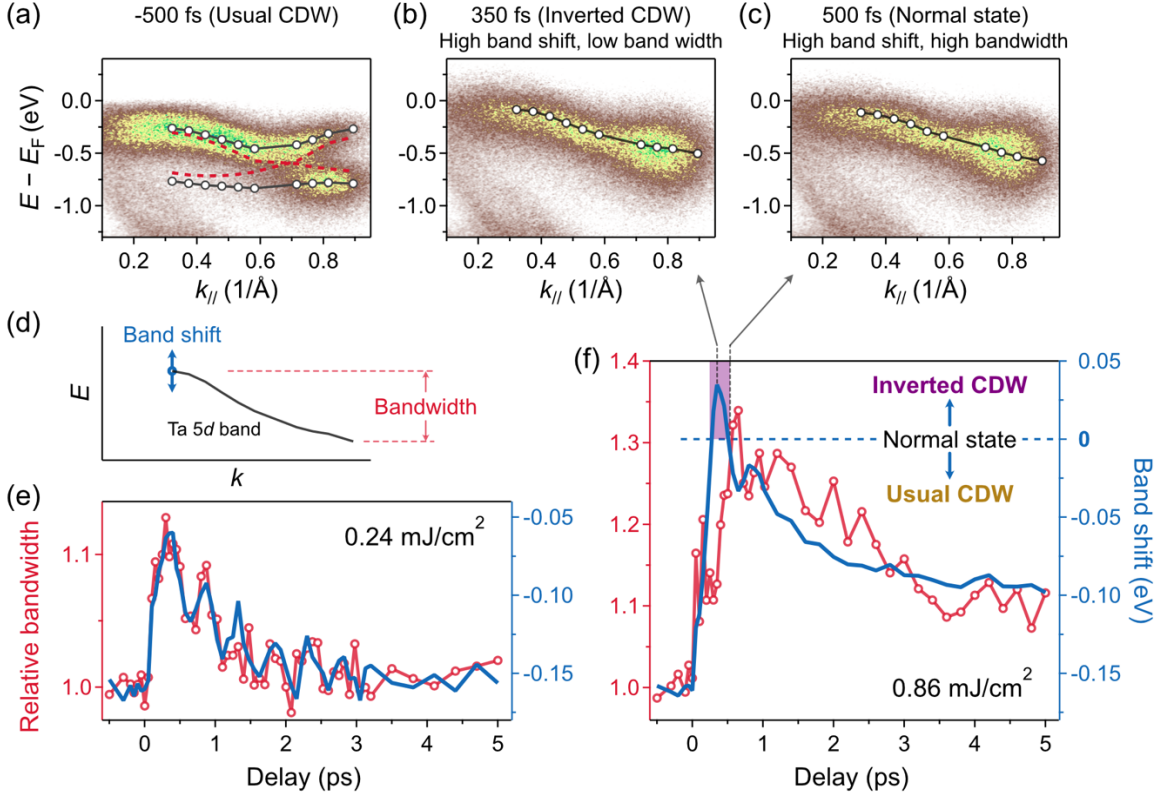

FIG. 2. Experimentally observed Ta 5*d* band dynamics. (a) ARPES intensity plot along the Γ-M direction before laser excitation at usual CDW state. The black and red curves represent the band dispersions before and after removing the hybridization gap. (b)–(c) Same as (a) but for the inverted CDW and normal states after laser excitation, respectively. (d) Definition of the band shift (energy shift at ~0.3 Å$^{-1}$ with respect to the normal state) and bandwidth (energy range of the dispersion). (e)–(f) Temporal evolution of the band shift (blue) and the relative bandwidth compared to that before laser excitation (red) at fluences of 0.24 mJ/cm$^2$ and 0.86 mJ/cm$^2$, respectively. Note that in (f) the horizontal blue dashed line aligned with a zero band shift corresponds to the normal state.

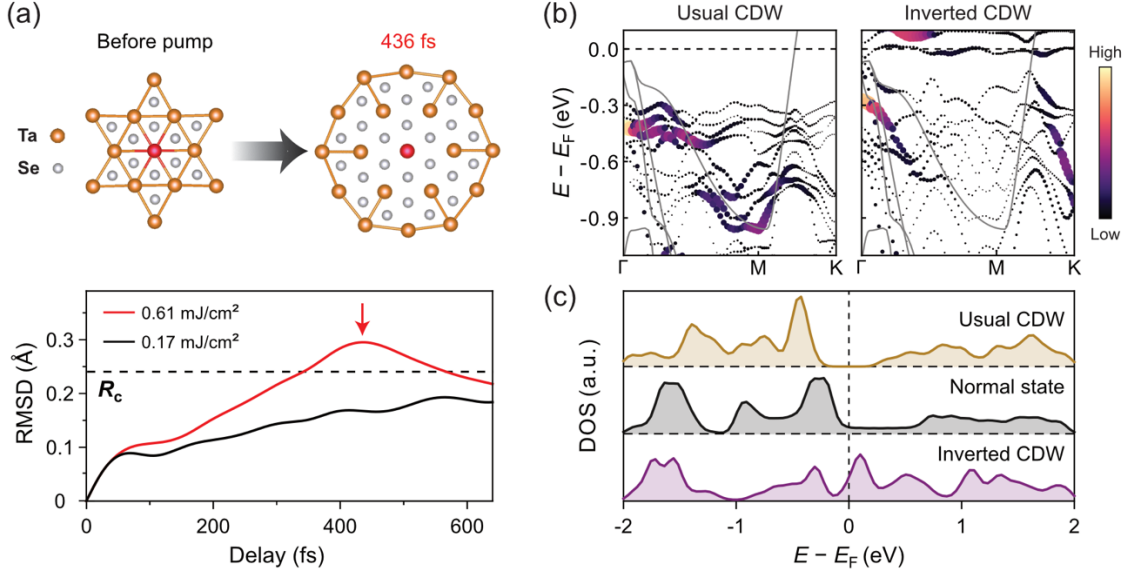

FIG. 3. TDDFT simulations of the structural and electronic dynamics. (a) The lower panel shows the atomic root-mean-square displacement evolution for two representative laser fluences. The dashed line indicates the critical value $R_c$ where the usual CDW order melts. The upper panel shows calculated structural patterns for the usual CDW and the inverted CDW (436 fs for a fluence of 0.61 mJ/cm$^2$) states. In both plots, the bonds between the Ta-Ta atoms that are shorter than 3.5 Å are colored to show the usual CDW (left) and the inverted CDW order (right). (b) Calculated unfolded band structures for the usual and inverted CDW states, respectively, where the spectral weight is reflected by the brightness. Here the magnitude of the periodic lattice distortion of the inverted CDW is set to be the same as the usual CDW. For comparison, the unreconstructed band structure of the normal state is indicated by the grey line in each panel. (c) Calculated in-plane density of states corresponding to (b) for different states, the one of the inverted CDW state is the highest near $E_F$.

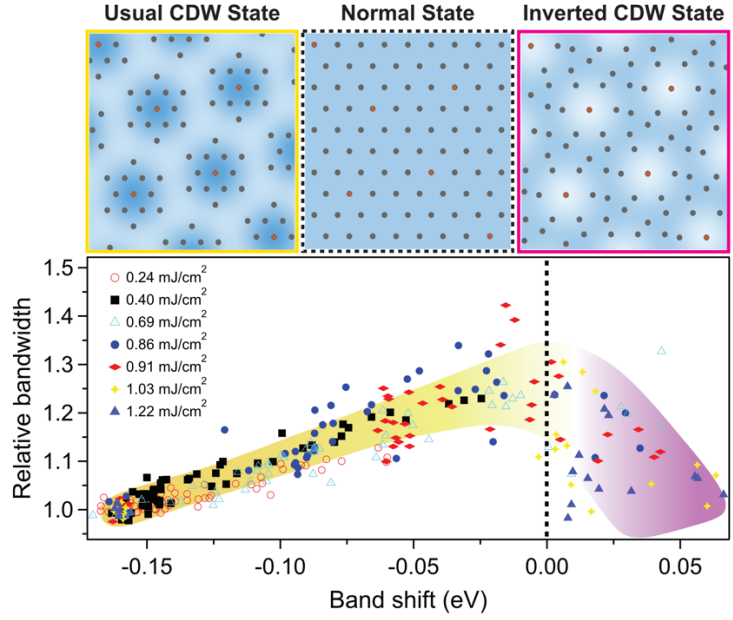

FIG. 4. The bandwidth as a function of band shift over a wide range of fluences and time delays. This universal relation can classify different states, with the lattice structures illustrated in upper panels. The yellow and purple shadows are the guide to the eye in the usual and inverted CDW states, respectively.

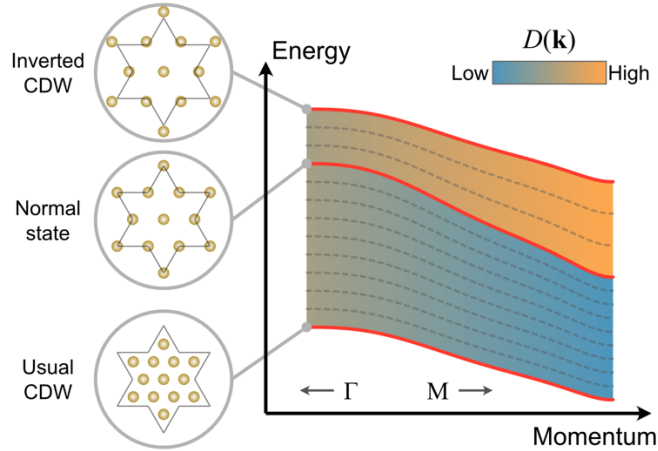

FIG. 5. The red solid lines and grey dashed lines represent the Ta $5d$ band dispersions corresponding to different periodic lattice distortions. The color scale of the shading represents the qualitative momentum dependence of the electron phonon coupling strength. Between the usual CDW state and the normal state, the deformation potential at the $\Gamma$ point is larger than that at the M point, while such gradient is reversed between the normal state and the inverted CDW state.